\documentclass[twocolumn,american,pre,twocolumns,showpacs]{revtex4-1}
\usepackage[T1]{fontenc}
\usepackage[latin9]{inputenc}
\usepackage{color}
\usepackage{babel}
\usepackage{textcomp}
\usepackage{amsmath}
\usepackage{amssymb}
\usepackage{graphicx}
\usepackage{esint}
\usepackage[unicode=true,pdfusetitle,
 bookmarks=true,bookmarksnumbered=false,bookmarksopen=false,
 breaklinks=false,pdfborder={0 0 0},backref=false,colorlinks=true]
 {hyperref}
\usepackage{breakurl}

\makeatletter

\newcommand{\lyxmathsym}[1]{\ifmmode\begingroup\def\b@ld{bold}
  \text{\ifx\math@version\b@ld\bfseries\fi#1}\endgroup\else#1\fi}

 \@ifundefined{textcolor}{}
 {%
   \definecolor{BLACK}{gray}{0}
   \definecolor{WHITE}{gray}{1}
   \definecolor{RED}{rgb}{1,0,0}
   \definecolor{GREEN}{rgb}{0,1,0}
   \definecolor{BLUE}{rgb}{0,0,1}
   \definecolor{CYAN}{cmyk}{1,0,0,0}
   \definecolor{MAGENTA}{cmyk}{0,1,0,0}
   \definecolor{YELLOW}{cmyk}{0,0,1,0}
 }

\makeatother

\begin{document}

\title{Shaping wave patterns in reaction-diffusion systems}

\author{Jakob Löber}

\email{jakob@physik.tu-berlin.de}

\author{Steffen Martens}

\author{Harald Engel}

\address{Institut für Theoretische Physik, Hardenbergstraße 36, EW 7-1, Technische
Universität Berlin, 10623 Berlin, Germany}
\begin{abstract}
We present a method to control the two-dimensional shape of traveling
wave solutions to reaction-diffusion systems, as e.g. interfaces and
excitation pulses. Control signals that realize a pre-given wave shape
are determined analytically from nonlinear evolution equation for
isoconcentration lines as the perturbed nonlinear phase diffusion
equation or the perturbed linear eikonal equation.\\
While the control enforces a desired wave shape perpendicular to the
local propagation direction, the wave profile along the propagation
direction itself remains almost unaffected. Provided that the one-dimensional
wave profile of all state variables and its propagation velocity can
be measured experimentally, and the diffusion coefficients of the
reacting species are given, the new approach can be applied even if
the underlying nonlinear reaction kinetics are unknown.
\end{abstract}

\keywords{traveling waves, control, reactions-diffusion systems}

\pacs{82.40.Ck, 02.30.Yy, 82.40.Bj}

\maketitle

\section{\label{sec:Introduction}Introduction}

Complex wave patterns in reaction-diffusion systems can often be
understood as being assembled of simple ``building blocks'' as traveling
fronts and pulses. A one-dimensional solitary pulse in the FitzHugh-Nagumo
model can be considered as being built of two propagating interfaces
separating the excited from the refractory state \cite{pismen2006patterns}.
These interfaces are front solutions to a simpler reaction-diffusion
system. Similarly, many two-dimensional shapes can be approximated
as consisting of appropriately shifted one-dimensional pulse profiles.
A reduced description entirely neglects the explicit form and dynamics
of the pulse profile and rather describes the shape of the pattern
in terms of a curve outlining it. Several evolution equations for
this time-dependent curve, called equations of motion (EOM) throughout
this article, can be derived directly from the reaction-diffusion
system \cite{kuramoto1980instability,zykov1987simulation,kuramoto2003chemical,dierckx2011accurate}.\\
The control of patterns in reaction-diffusion system has received
the attention of many researchers in the past \cite{mikhailov2006control,vanag2008design}.
For example, different feedback control loops have been realized in
experiments with the photosensitive Belousov-Zhabotinsky (BZ) reaction
\cite{krug1990analysis} using feedback signals obtained from wave
activity measured at one or several detector points, along detector
lines, or in a spatially extended control domain including global
feedback control \cite{PhysRevLett.92.018304,zykov2004feedback,schlesner2008efficient}.
Furthermore, feedback-mediated control loops can be employed in order
to stabilize unstable patterns, such as unstable traveling wave segments
and spots. Two feedback loops were used to guide unstable wave segments
in the BZ reaction along pre-given trajectories \cite{sakurai2002design,mihaliuk2002feedback}.
An open loop control was successfully deployed in dragging traveling
chemical pulses of adsorbed CO during heterogeneous catalysis on platinum
single crystal surfaces \cite{wolff2003gentle}. In these experiments,
the pulse velocity was controlled by a laser beam creating a movable
localized temperature heterogeneity on an addressable catalyst surface,
resulting in a V-shaped pattern \cite{wolff2001spatiotemporal,wolff2003wave}.
Dragging a one-dimensional chemical front or phase interface to a
new position by anchoring it to a movable parameter heterogeneity,
was studied theoretically in \cite{nistazakis2002targeted,malomed2002pulled,kevrekidis2004dragging}.\\
Recently we developed a method to control the position over time of
one-dimensional traveling waves in reaction-diffusion systems by spatio-temporal
forcing \cite{loeber2013}. We utilized an ordinary differential equation
for the wave's position over time in response to an external perturbation.
Using this equation, we formulated an inverse problem for the control
signal enforcing the traveling wave to follow a desired protocol of
motion. We demonstrated by example that the analytically obtained
control function is close to the numerical solution of an appropriately
formulated optimal control algorithm \cite{buchholz2013on,troltzsch2010optimal,jorge1999numerical}.
Furthermore, we identified the mechanism leading to a successful position
control and thereby proved stability of position control with respect
to perturbations of the initial position on the reduced level of the
one-dimensional EOM \cite{loeber2013Stability}.\\
In this article we extend position control to shape control of two-dimensional
wave patterns. In Sec. \ref{sec:EquationOfMotionForTravelingWaves},
EOMs for two-dimensional traveling waves under perturbations are motivated.
We demonstrate how these equations can be utilized to control the
shape of a nearly planar traveling wave in Sec. \ref{sec:ShapingATraveling},
and of more complex wave patterns in Sec. \ref{sec:ShapingPatterns}.
In Sec. \ref{sec:PositionControlInFiniteDomains} we discuss which
additional control terms have to act on the boundary of a finite domain
to guide traveling waves from the outside to the inside of a domain
or vice versa. We end with conclusions in Sec. \ref{sec:Conclusions}.

\section{\label{sec:EquationOfMotionForTravelingWaves}Equation of motion
for two-dimensional traveling waves}

We consider a perturbed reaction-diffusion (RD) system for the vector
of $n$ species ${\mathbf{u}=\mathbf{u}\left(\mathbf{r},t\right)=\left(u_{1}\left(\mathbf{r},t\right),\dots,u_{n}\left(\mathbf{r},t\right)\right)^{T}}$
in a two-dimensional medium with position vector ${\mathbf{r}=\left(x,y\right)^{T}}$,
\begin{eqnarray}
\partial_{t}\mathbf{u} & =\mathcal{D}\triangle\mathbf{u}+\mathbf{R}\left(\mathbf{u}\right)+\epsilon\mathcal{G}\left(\mathbf{u}\right)\mathbf{f}\left(\mathbf{r},t\right).\label{eq:ControlledRDS}
\end{eqnarray}
Here, $\mathcal{D}$ is a diagonal matrix of constant diffusion coefficients
and $\triangle$ denotes the Laplacian operator. The spatiotemporal
perturbations $\mathbf{f}$ are coupled by a (possibly $\mathbf{u}$-dependent)
matrix $\mathcal{G}$ into the system, and $\mathbf{R}$ describes\textcolor{red}{{}
}a nonlinear reaction kinetics. The small parameter $\epsilon$ ensures
that the perturbation couples weakly to the system and can be used
for a perturbation expansion \cite{loeber2013}. The medium is assumed
to be isotropic, infinitely extended in the $x$-direction, and finite
with domain size $L_{y}$ in the $y$-direction. Homogeneous Neumann
boundary conditions are applied in the $y$-direction, 
\begin{eqnarray}
\partial_{y}\mathbf{u}\left(x,0,t\right) & =\partial_{y}\mathbf{u}\left(x,L_{y},t\right)=0.\label{eq:NeumannBoundary}
\end{eqnarray}
Solutions of interest of the unperturbed ($\epsilon=0$) RD system
Eq. \eqref{eq:ControlledRDS} are planar traveling waves $\mathbf{U}_{c}$.
We fix the propagation direction of the unperturbed wave to be the
$x$-direction such that $\mathbf{U}_{c}\left(\xi\right)$ is a stationary
solution in a frame of reference $\xi=x-ct$ comoving with velocity
$c$, 
\begin{eqnarray}
0 & =\mathcal{D}\partial_{\xi}^{2}\mathbf{U}_{c}\left(\xi\right)+c\partial_{\xi}\mathbf{U}_{c}\left(\xi\right)+\mathbf{R}\left(\mathbf{U}_{c}\left(\xi\right)\right).\label{eq:ProfileEquation}
\end{eqnarray}
Upon a perturbation expansion of Eq. \eqref{eq:ControlledRDS} around
the traveling wave solution $\mathbf{U}_{c}$, the linear operator
\begin{eqnarray}
\mathcal{L} & =\mathcal{D}\partial_{\xi}^{2}+c\partial_{\xi}+\mathbb{D}\mathbf{R}\left(\mathbf{U}_{c}\left(\xi\right)\right),\label{eq:LinearOperatorL}
\end{eqnarray}
arises \cite{keener1986geometrical,zykov1987simulation}. The derivative
of the unperturbed wave profile $\mathbf{U}_{c}\left(\xi\right)$
with respect to $\xi$, $\mathbf{W}\left(\xi\right)=\mathbf{U}_{c}'\left(\xi\right)$,
is an eigenfunction of the operator $\mathcal{L}$ to eigenvalue zero,

\begin{eqnarray}
\mathcal{L}\mathbf{U}_{c}'\left(\xi\right) & =0,
\end{eqnarray}
known as the Goldstone mode. The existence of the Goldstone mode corresponds
to the invariance of the unperturbed RD system with respect to spatial
translations in the $x$-direction. We assume $\mathbf{U}_{c}\left(\xi\right)$
to be a \textit{stable} traveling wave solution such that the zero
eigenvalue of $\mathcal{L}$, $\lambda_{0}=0$, is the eigenvalue
with largest real part. Furthermore, we suppose the existence of a
spectral gap, i.e., the eigenvalue $\lambda_{1}$ of $\mathcal{L}$
with next largest real part is separated by a finite distance from
the imaginary axis.\\
For what follows we also need the adjoint Goldstone mode or response
function $\mathbf{W}^{\dagger}\left(x\right)$ \cite{biktasheva1998localized}
defined as the eigenfunction to eigenvalue $0$ of the adjoint operator
$\mathcal{L}^{\dagger}$ of $\mathcal{L}$,

\begin{eqnarray}
\mathcal{L}^{\dagger} & =\mathcal{D}\partial_{\xi}^{2}-c\partial_{\xi}+\mathbb{D}\mathbf{R}\left(\mathbf{U}_{c}\left(\xi\right)\right)^{T},\label{eq:AdjointLinearOperatorLdg}
\end{eqnarray}
\begin{eqnarray}
\mathcal{L}^{\dagger}\mathbf{W}^{\dagger} & =0.\label{eq:ResponseFunctionEquation}
\end{eqnarray}
For single component RD systems with scalar diffusion coefficient
$\mathcal{D}=D$, a general expression for the response function in
terms of the traveling wave profile $U_{c}$ reads
\begin{eqnarray}
W^{\dagger}\left(\xi\right) & =e^{c\xi/D}U_{c}'\left(\xi\right),
\end{eqnarray}
while in the multi-component case no general expression is known.\\
Perturbations can deform both the profile $\mathbf{U}_{c}$ and the
shape of a traveling plane wave solution. We neglect deformations
of the wave profile $\mathbf{U}_{c}$ and describe the deviations
from the planar wave shape by a time-dependent curve ${\boldsymbol{\gamma}\left(s,t\right)=\left(\gamma_{x}\left(s,t\right),\gamma_{y}\left(s,t\right)\right)^{T}}$
tracing out a chosen isoconcentration line parametrized by $s$. This
curve\textcolor{red}{{} }specifies the position of a traveling wave
in two spatial dimensions. For a monotonically decreasing front solution,
we define its position as the point of steepest slope, while the position
of a pulse solution is given by the point of maximum amplitude of
an arbitrary component. For an unperturbed plane wave solution propagating
with velocity $c$ in the $x$-direction, the curve $\boldsymbol{\gamma}$
is a straight line given by
\begin{eqnarray}
\boldsymbol{\gamma}\left(y,t\right) & = & \left(\begin{array}{c}
ct\\
y
\end{array}\right),
\end{eqnarray}
where $y$ denotes\textcolor{red}{{} }the Cartesian coordinate transversal
to the propagation direction. A local velocity for each point of the
curve can be defined as (throughout the article, the time derivative
is indicated by the dot while the prime denotes the derivative with
respect to the spatial variable)
\begin{eqnarray}
\mathbf{v}\left(s,t\right) & =\dot{\boldsymbol{\gamma}}\left(s,t\right).
\end{eqnarray}
For a straight line, this yields a velocity $\mathbf{v}\left(y,t\right)=c\,\mathbf{e}_{x}$
which is constant along the curve and equals the overall velocity
$c$ of the plane wave.\\
Using multiple scale perturbation theory, an evolution equation for
$\boldsymbol{\gamma}\left(s,t\right)$ can be derived from the perturbed
RD system Eq. \eqref{eq:ControlledRDS}. We suppose the following
parametrization for $\boldsymbol{\gamma}$ 
\begin{eqnarray}
\boldsymbol{\gamma}\left(y,t\right) & = & \left(\begin{array}{c}
\phi\left(y,t\right)\\
y
\end{array}\right),\label{eq:NonlinearPhaseDiffusionEquationParametrization}
\end{eqnarray}
to obtain an equation for the $x$-component of the position $\phi\left(y,t\right)$.
The ansatz 
\begin{eqnarray}
\phi\left(y,t\right) & = & ct+\Phi\left(Y,T\right).
\end{eqnarray}
assumes a fast dynamics on the time scale $t$ which corresponds to
the ordinary one-dimensional propagation of the unperturbed plane
wave with velocity $c$. In a frame comoving with velocity $c$, all
variations, denoted by $\Phi$, are assumed to be slow. Consequently,
$\Phi$ depends on the slow time scale $T=\epsilon t$. Furthermore,
it is assumed that the curve $\boldsymbol{\gamma}$ varies only weakly
along the transversal direction, i.e., $\Phi$ depends solely on the
stretched spatial coordinate $Y=\epsilon^{1/2}y$ \cite{kuramoto1980instability}.
The small parameter $\epsilon$ fixes the time and space scale on
which the curve $\boldsymbol{\gamma}$ changes in the comoving frame
as well as the amplitude of the perturbation in Eq. \eqref{eq:ControlledRDS}.
Because of the presumed existence of a spectral gap of $\mathcal{L}$,
deformations of the pulse profile in response to the perturbation
decay fast and can be neglected \cite{loeber2013}. These assumptions
yield a closed EOM for $\phi\left(y,t\right)$ \cite{kuramoto1980instability},

\begin{align}
\dot{\phi} & =c+\frac{c}{2}\left(\phi'\right)^{2}+\alpha\phi''\nonumber \\
 & -\frac{\epsilon}{K_{c}}\intop_{-\infty}^{\infty}dx\mathbf{W}^{\dagger T}\left(x\right)\mathcal{G}\left(\mathbf{U}_{c}\left(x\right)\right)\mathbf{f}\left(\mathbf{r}+\phi\mathbf{e}_{x},t\right),\label{eq:PerturbedPhaseDiffusionEquation}
\end{align}
with the constants 
\begin{eqnarray}
K_{c} & =\left\langle \mathbf{W}^{\dagger},\mathbf{U}_{c}'\right\rangle = & \intop_{-\infty}^{\infty}dx\mathbf{W}^{\dagger T}\left(x\right)\mathbf{U}_{c}'\left(x\right),\\
\alpha & =\frac{\left\langle \mathbf{W}^{\dagger},\mathcal{D}\mathbf{U}_{c}'\right\rangle }{\left\langle \mathbf{W}^{\dagger},\mathbf{U}_{c}'\right\rangle }= & \frac{1}{K_{c}}\intop_{-\infty}^{\infty}dx\mathbf{W}^{\dagger T}\left(x\right)\mathcal{D}\mathbf{U}_{c}'\left(x\right).\label{eq:Alpha}
\end{eqnarray}
and initial condition 
\begin{eqnarray}
\phi\left(y,t_{0}\right) & = & \phi_{0}\left(y\right).
\end{eqnarray}
A detailed derivation of Eq. \eqref{eq:PerturbedPhaseDiffusionEquation}
is given in Appendix \ref{sec:DerivationPNPDE}. For simplicity, we
assume $\phi$ to be a single valued function of $y$. This excludes
any overhangs and in particular closed curves $\boldsymbol{\gamma}$.
Therefore an isoconcentration line described by $\boldsymbol{\gamma}$
has to start and end at the domain boundaries. A Neumann boundary
condition, Eq. \eqref{eq:NeumannBoundary}, for the RD system \eqref{eq:ControlledRDS}
implies a right angle between any isoconcentration line and the domain
boundary and translates to a Neumann boundary for $\phi$,
\begin{eqnarray}
\partial_{y}\phi\left(0,t\right) & = & \partial_{y}\phi\left(L_{y},t\right)=0.
\end{eqnarray}
The unperturbed ($\epsilon=0$) version of Eq. \eqref{eq:PerturbedPhaseDiffusionEquation}
is known as the nonlinear phase diffusion equation \cite{kuramoto1980instability,kuramoto2003chemical}.
This nonlinear PDE can be transformed to the linear diffusion equation
via the Cole-Hopf transform. The nonlinear term $\sim\left(\phi'\right)^{2}$
in Eq. \eqref{eq:PerturbedPhaseDiffusionEquation} has a purely geometric
origin while the second order derivative $\sim\phi''$ describes relaxation
of the curve to a straight line with a surface tension $\alpha$,
Eq. \eqref{eq:Alpha}. Note that for a single component RD system
with scalar diffusion coefficient $\mathcal{D}=D$ follows $\alpha=D>0$.
Negative values of $\alpha$ can occur in unperturbed RD systems of
activator-inhibitor type if the inhibitor diffuses much faster than
the activator \cite{zykov1998waveinstabilities,horvath1993instabilities}.
For $\alpha<0$ plane wave solutions become unstable, undergoing a
transversal wave instability. Introducing a fourth order derivative
$\sim\partial_{y}^{4}\phi$ in the unperturbed ($\epsilon=0$) Eq.
\eqref{eq:PerturbedPhaseDiffusionEquation} leads to the Kuramoto-Sivashinsky
equation which can describe patterns arising beyond the onset of transversal
instabilities \cite{kuramoto1980instability,malevanets1995biscale}.
A variation of Eq. \eqref{eq:PerturbedPhaseDiffusionEquation} with
a spatially distributed Gaussian white noise term instead of the perturbation
is known as the Kardar-Parisi-Zhang equation \cite{kardar1986dynamic}.\\
For plane waves $\phi$ does not depend on $y$ and the perturbed
nonlinear phase diffusion equation \eqref{eq:PerturbedPhaseDiffusionEquation}
reduces to the one-dimensional case 
\begin{align}
\dot{\phi}\left(t\right) & =c-\frac{\epsilon}{K_{c}}\intop_{-\infty}^{\infty}dx\mathbf{W}^{\dagger T}\left(x\right)\mathcal{G}\left(\mathbf{U}_{c}\left(x\right)\right)\mathbf{f}\left(x+\phi\left(t\right),t\right),\label{eq:1DimEquationOfMotion}
\end{align}
which has been intensively studied in \cite{schimanskygeier1983effect,engel1985noise,engel1987interaction,kulka1995influence,bode1997front,alonso2010wave,loeber2012front}.

\section{\label{sec:ShapingATraveling}Shaping a plane wave}

Usually the EOM \eqref{eq:PerturbedPhaseDiffusionEquation} is seen
as a PDE for the wave shape $\phi\left(y,t\right)$. However, here
we utilize Eq. \eqref{eq:PerturbedPhaseDiffusionEquation} to formulate
the inverse problem. Namely, we are looking for the control signal
$\mathbf{f}$ given as a solution to the integral equation \eqref{eq:PerturbedPhaseDiffusionEquation}
for a prescribed function $\phi\left(y,t\right)$ outlining the shape
of the pattern to be enforced \cite{loeber2013}.\\
Many inverse problems are mathematically ill posed as they do not
yield a unique solution. This holds in our case, too. Therefore, we
have to rely on some physical insight to pick a meaningful solution.
We start from a fairly general solution involving some arbitrary functions
which are then fixed in a second step. Below, we set $\epsilon=1$,
assuming the control signal $\mathbf{f}$ is sufficiently small in
amplitude. Furthermore, we suppose that the wave moves unperturbed
until the initial time $t=t{}_{0}$ upon which the control is switched
on. The initial shape of the wave is not necessarily a plane wave
$\phi\left(y,t_{0}\right)=\phi_{0}\left(y\right)$.\\
We rearrange Eq. \eqref{eq:PerturbedPhaseDiffusionEquation} in the
form

\begin{eqnarray}
K_{c}\left(c-\dot{\phi}+\frac{c}{2}\left(\phi'\right)^{2}+\alpha\phi''\right) & =\nonumber \\
\intop_{-\infty}^{\infty}dx\mathbf{W}^{\dagger T}\left(x\right)\mathcal{G}\left(\mathbf{U}_{c}\left(x\right)\right)\mathbf{f}\left(\mathbf{r}+\phi\mathbf{e}_{x},t\right).\label{eq:NPDE1}
\end{eqnarray}
The left hand side is a sum over four terms, thus, to be as general
as possible, we assume a superposition of four independent contributions
to the control term $\mathbf{f}$ according to
\begin{align}
\mathbf{f}\left(\mathbf{r},t\right) & =\mathcal{G}^{-1}\left(\mathbf{U}_{c}\left(x-\phi\right)\right)\sum_{i=1}^{4}\mathcal{A}_{i}\left(y,t\right)\mathbf{h}_{i}\left(x-\phi\right).\label{eq:ControlAnsatz}
\end{align}
To eliminate the coupling matrix $\mathcal{G}$, all terms are multiplied
by the inverse $\mathcal{G}^{-1}$. Because all terms depending on
the $x$-coordinate are evaluated at $x-\phi$, $\phi$ can be eliminated
by a simple substitution in the integral on the right hand side of
Eq. \eqref{eq:NPDE1}. The terms $\mathcal{G}^{-1}\left(\mathbf{U}_{c}\left(x-\phi\right)\right)\mathbf{h}_{i}\left(x-\phi\right)$
are constant in the comoving frame of reference of the controlled
traveling wave while the control amplitude is determined by the terms
$\mathcal{A}_{i}\left(y,t\right)$. We are left with
\begin{align}
K_{c}\left(c-\dot{\phi}+\frac{c}{2}\left(\phi'\right)^{2}+\alpha\phi''\right) & =\nonumber \\
\sum_{i=1}^{4}\mathcal{A}_{i}\left(y,t\right)\left\langle \mathbf{W}^{\dagger}\left(x\right),\mathbf{h}_{i}\left(x\right)\right\rangle .
\end{align}
Every term on the right hand side cancels exactly one term on the
left hand side if we set
\begin{eqnarray}
\mathcal{A}_{1}\left(y,t\right) & =c\frac{K_{c}}{G_{1}}, & \mathcal{A}_{2}\left(y,t\right)=-\dot{\phi}\frac{K_{c}}{G_{2}},\nonumber \\
\mathcal{A}_{3}\left(y,t\right) & =\frac{c}{2}\left(\phi'\right)^{2}\frac{K_{c}}{G_{3}},\; & \mathcal{A}_{4}\left(y,t\right)=\alpha\phi''\frac{K_{c}}{G_{3}},\label{eq:AAnsatz}
\end{eqnarray}
with
\begin{eqnarray}
G_{i} & =\left\langle \mathbf{W}^{\dagger}\left(x\right),\mathbf{h}_{i}\left(x\right)\right\rangle = & \intop_{-\infty}^{\infty}dx\mathbf{W}^{\dagger T}\left(x\right)\mathbf{h}_{i}\left(x\right).\label{eq:Gi}
\end{eqnarray}
Equation \eqref{eq:ControlAnsatz} together with Eq. \eqref{eq:AAnsatz}
and Eq. \eqref{eq:Gi} constitutes the most general solution for the
control function $\mathbf{f}$. This solution contains four arbitrary
vector-valued functions $\mathbf{h}_{i}\left(x\right)$. 
\begin{figure}
\begin{center}\includegraphics{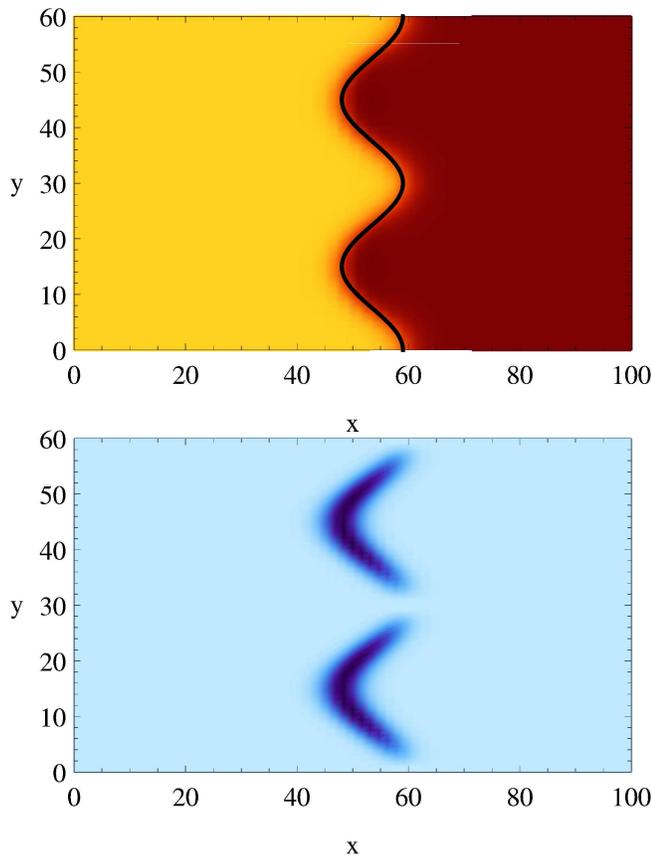}\end{center}\caption{\label{fig:ShapingASchloeglFront}(Color online) Top: Shaping a Schlögl
front according to a protocol $\phi\left(y,t\right)$ (black line).
Yellow (red) corresponds to a high (low) value of $u$. Bottom: Corresponding
control function $f\left(\mathbf{r},t\right)$. Blue (black) corresponds
to a high (low) value of $f$. See movie in the Supplemental Material
\cite{supplement}.}
\end{figure}
Because a perturbation proportional to the Goldstone mode, $\mathbf{f}\sim\mathbf{U}_{c}'$,
shifts the traveling wave as a whole \cite{loeber2013}, we choose
\begin{eqnarray}
\mathbf{h}_{i}\left(x\right) & =\mathbf{U}_{c}'\left(x\right),\, i\in\left\{ 1,\dots,4\right\} 
\end{eqnarray}
and get
\begin{align}
\mathbf{f} & =\left(c-\dot{\phi}+\frac{c}{2}\phi'^{2}+\alpha\phi''\right)\mathcal{G}^{-1}\left(\mathbf{U}_{c}\left(x-\phi\right)\right)\mathbf{U}_{c}'\left(x-\phi\right).\label{eq:TwoDimControl}
\end{align}
Note that the constants $G_{i}$ involving the response function $\mathbf{W}^{\dagger}\left(x\right)$
cancel out because $G_{c}=K_{c}$ for this choice. This is of great
advantage because, in general, the response function can only be determined
numerically from the one-dimensional RD equations, or by repeated
measurements of traveling waves in a well defined experimental setup.
However, the coefficient $\alpha$, Eq. \eqref{eq:Alpha}, still depends
on the response function. To get rid of $\alpha$, we choose a slightly
different solution with
\begin{eqnarray}
\mathbf{h}_{i}\left(x\right) & = & \mathbf{U}_{c}'\left(x\right),\, i\in\left\{ 1,\dots,3\right\} ,\\
\mathbf{h}_{4}\left(x\right) & = & \mathcal{D}\mathbf{U}_{c}'\left(x\right),
\end{eqnarray}
and find the following solution for the control
\begin{align}
\mathbf{f} & =\mathcal{G}^{-1}\left(\mathbf{U}_{c}\left(x-\phi\right)\right)\left(\left(c-\dot{\phi}+\frac{c}{2}\phi'^{2}\right)\mathbf{U}_{c}'\left(x-\phi\right)\right.\nonumber \\
 & \left.\vphantom{\left(c-\dot{\phi}+\frac{c}{2}\phi'^{2}\right)\mathbf{U}_{c}'\left(x-\phi\right)}+\phi''\mathcal{D}\mathbf{U}_{c}'\left(x\right)\right).\label{eq:TwoDimControl2}
\end{align}
Instead of the (unknown) coefficient $\alpha$, Eq. \eqref{eq:TwoDimControl2}
contains the matrix of diffusion coefficients $\mathcal{D}$. Note
that for the single component case with scalar diffusion coefficient
$\mathcal{D}=D$ we obtain $\alpha=D$ and Eq. \eqref{eq:TwoDimControl2}
is equivalent to Eq. \eqref{eq:TwoDimControl}.\\
In the following, we apply the shape control Eq. \eqref{eq:TwoDimControl}
to the front solution of the Schlögl model \cite{Schlogl1972crm},
\begin{eqnarray}
\partial_{t}u & =\triangle u-u\left(u-a\right)\left(u-1\right).\label{eq:SchloeglModel}
\end{eqnarray}
The Schlögl model is a simple example of a single component RD system
exhibiting bistability. Initially, Eq. \eqref{eq:SchloeglModel} has
been discussed in 1938 by Zeldovich and Frank-Kamenetsky in connection
with flame propagation \cite{zeldovich1938theory}. The one-dimensional
front solution connects the homogeneous steady state $u=1$ for $x\rightarrow-\infty$
and $u=0$ for $x\rightarrow\infty$. The front profile is known analytically
\cite{mikhailov1990foundations} 
\begin{eqnarray}
U_{c}\left(\xi\right) & =1/\left(1+\exp\left(\xi/\sqrt{2}\right)\right)
\end{eqnarray}
 and the associated propagation velocity is given by
\begin{eqnarray}
c & =\frac{1}{\sqrt{2}}\left(1-2a\right).
\end{eqnarray}
As an example, we choose an additive control, i.e., a constant coupling
function $\mathcal{G}\left(u\right)=1$, in the control signal Eq.
\eqref{eq:TwoDimControl}.\\
Starting from a plane front traveling in the $x$-direction at $t=t_{0}$,
we want the control to enforce a transition to a sinusoidally shaped
stationary front at time $t=t_{1}$. The corresponding protocol for
$t_{0}\leq t\leq t_{1}$ is
\begin{eqnarray}
\phi\left(y,t\right) & =\phi_{0}+A\cos\left(4\pi\dfrac{y}{L_{y}}\right)\sin\left(\dfrac{\pi}{2}\dfrac{t-t_{0}}{t_{1}-t_{0}}\right).
\end{eqnarray}
In numerical simulations, we set $L_{y}=60$ for the domain size in
the $y$-direction, $t_{1}-t_{0}=300/4$ for the time span and $A=50/\left(2\pi\right)$
for the amplitude. Starting at time $t_{1}$, the protocol is frozen
as $\phi\left(y,t_{1}\right)$ to maintain a sinusoidally modulated
stationary front; cf. the movie in \cite{supplement}. Fig. \ref{fig:ShapingASchloeglFront}
displays a snapshot showing the time evolution of a Schlögl front
under the control. The shape of the front, given by the points of
steepest slope of the front profile, closely follows the protocol
(black solid line), see Fig. \ref{fig:ShapingASchloeglFront} top.
The corresponding control function attains its largest amplitude at
the points where the modulated shape deviates most from a plane wave,
see Fig. \ref{fig:ShapingASchloeglFront} bottom.

\section{\label{sec:ShapingPatterns}Shaping arbitrary wave patterns}

The nonlinear phase diffusion equation \eqref{eq:PerturbedPhaseDiffusionEquation}
arises through a perturbation expansion around a one-dimensional traveling
wave propagating in the $x$-direction. Naturally, we expect Eq. \eqref{eq:PerturbedPhaseDiffusionEquation}
to fail if the local propagation direction of the perturbed wave is
very different from the $x$-direction. For example, a plane wave
traveling in the $y$-direction cannot be a solution to Eq. \eqref{eq:PerturbedPhaseDiffusionEquation}.
Furthermore, we also want to describe and control wave patterns outlined
by a closed curve $\boldsymbol{\gamma}\left(s,t\right)$. A generalized
EOM, known as the linear eikonal equation, accounts for such cases.
To derive the latter, a local coordinate system in terms of the coordinates
$s$ and $\rho$ is constructed,
\begin{figure}
\begin{center}\includegraphics[scale=0.7]{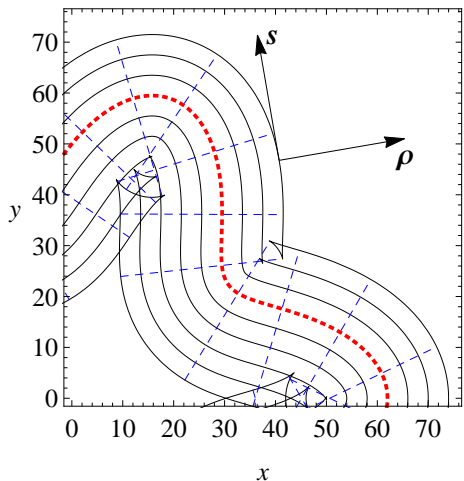}\end{center}

\caption{\label{fig:CoordinateSystem}(Color online) Time-dependent local coordinate
system $\left(\rho,s\right)$ used for the derivation of the linear
eikonal equation. Dashed blue lines are contours with $s=\text{const.}$,
black solid lines are contours with $\rho=\text{const}$. The red
dotted line denotes the curve $\boldsymbol{\gamma}\left(s,t\right)$
which corresponds to the contour with $\rho=0$. $\boldsymbol{\gamma}\left(s,t\right)$
defines the position of the traveling wave in two spatial dimensions
and outlines the desired shape of the controlled wave pattern.}
\end{figure}
\begin{align}
\mathbf{r} & =\left(\begin{array}{c}
x\left(\rho,s,t\right)\\
y\left(\rho,s,t\right)
\end{array}\right)=\boldsymbol{\mathbf{\chi}}\left(\rho,s,t\right)=\boldsymbol{\gamma}\left(s,t\right)+\rho\mathbf{n}\left(s,t\right).\label{eq:CoordinateTransform}
\end{align}
Here $\mathbf{n}\left(s,t\right)$ is the normal vector of $\boldsymbol{\gamma}$,
\begin{eqnarray}
\mathbf{n}\left(s,t\right) & = & \frac{\boldsymbol{\gamma}''\left(s,t\right)}{\sqrt{\boldsymbol{\gamma}''\left(s,t\right)\cdot\boldsymbol{\gamma}''\left(s,t\right)}}.
\end{eqnarray}
In contrast to the nonlinear phase diffusion equation, which assumes
a fixed propagation direction of the unperturbed traveling wave, here
propagation is along the normal direction which is allowed to vary
in time and space. The leading order solution to the RD system expressed
in the new coordinates is the one-dimensional traveling wave solution
$\mathbf{U}_{c}\left(\rho\right)$. See Fig. \ref{fig:CoordinateSystem}
for a visualization of the coordinate system Eq. \eqref{eq:CoordinateTransform}.\\
The normal velocity $c_{n}$ along the curve $\boldsymbol{\gamma}$
is defined as
\begin{eqnarray}
c_{n}\left(s,t\right) & = & \mathbf{n}\left(s,t\right)\cdot\dot{\boldsymbol{\gamma}}\left(s,t\right),
\end{eqnarray}
while its curvature $\kappa\left(s,t\right)$ is given by
\begin{eqnarray}
\kappa & = & \frac{\gamma_{x}'\gamma_{y}''-\gamma_{y}'\gamma_{x}''}{\left(\left(\gamma_{x}'\right)^{2}+\left(\gamma_{y}'\right)^{2}\right)^{3/2}}.
\end{eqnarray}
The linear eikonal equation relates the normal velocity $c_{n}$ along
the curve $\boldsymbol{\gamma}$ to its local curvature $\kappa$,

\begin{eqnarray}
c_{n}\left(s,t\right) & = & c-\alpha\kappa\left(s,t\right).\label{eq:LinearEikonalEquation}
\end{eqnarray}
Recently, Dierckx et al. \cite{dierckx2011accurate} derived higher
order corrections to Eq. \eqref{eq:LinearEikonalEquation} and generalized
it to anisotropic media. In particular, they showed that for isotropic
media the coefficient $\alpha$ is indeed given by expression \eqref{eq:Alpha}.\\
Equation \eqref{eq:LinearEikonalEquation} is a coordinate-free expression
for the evolution law of $\boldsymbol{\gamma}\left(s,t\right)$. To
account for the effect of a spatio-temporal perturbation $\mathbf{f}\left(\mathbf{r},t\right)$,
we suggest the following generalization of Eq. \eqref{eq:LinearEikonalEquation},
\begin{align}
c_{n}\left(s,t\right) & =c-\alpha\kappa\left(s,t\right)\label{eq:PerturbedLinearEikonalEquation}\\
 & -\frac{\epsilon}{K_{c}}\intop_{-\infty}^{\infty}d\rho\mathbf{W}^{\dagger T}\left(\rho\right)\mathcal{G}\left(\mathbf{U}_{c}\left(\rho\right)\right)\mathbf{f}\left(\boldsymbol{\mathbf{\chi}}\left(\rho,s,t\right),t\right).\nonumber 
\end{align}
The integration is performed over the coordinate $\rho$ longitudinal
to the local propagation direction. The perturbative term is a direct
generalization of the corresponding term in the perturbed nonlinear
phase diffusion equation Eq. \eqref{eq:PerturbedPhaseDiffusionEquation}.
Supposing the same time and space scale separation as in Sec. \ref{sec:EquationOfMotionForTravelingWaves}
and Appendix \ref{sec:DerivationPNPDE}, one obtains the perturbed
phase diffusion equation \eqref{eq:PerturbedPhaseDiffusionEquation}
from the perturbed linear eikonal equation \eqref{eq:PerturbedLinearEikonalEquation}
(see Appendix \ref{sec:FromThePLEEToThePNPDE} for a detailed derivation).
Furthermore, the derivation reveals that the curvature coefficient
$\alpha$ in Eq. \eqref{eq:PerturbedLinearEikonalEquation} indeed
equals the coefficient in front of the 2nd order derivative in Eq.
\eqref{eq:PerturbedPhaseDiffusionEquation}.\\
We employ the perturbed linear eikonal equation \eqref{eq:PerturbedLinearEikonalEquation}
to derive a control signal enforcing a desired shape while simultaneously
preserving the one-dimensional wave profile $\mathbf{U}_{c}\left(\rho\right)$
along the coordinate $\rho$. One possible solution of the integral
equation \eqref{eq:PerturbedLinearEikonalEquation} for the control
function is given by
\begin{align}
\mathbf{f}\left(\boldsymbol{\mathbf{\chi}}\left(\rho,s,t\right),t\right) & =\left(c-c_{n}-\alpha\kappa\right)\mathcal{G}^{-1}\left(\mathbf{U}_{c}\left(\rho\right)\right)\mathbf{U}_{c}'\left(\rho\right).\label{eq:ControlEikonal0}
\end{align}
Similar as in case of the nonlinear phase diffusion equation, we can
also find another solution for $\mathbf{f}$ which does not involve
$\alpha$,
\begin{align}
\mathbf{f}\left(\boldsymbol{\mathbf{\chi}}\left(\rho,s,t\right),t\right) & =\mathcal{G}^{-1}\left(\mathbf{U}_{c}\left(\rho\right)\right)\left(c-c_{n}-\mathcal{D}\kappa\right)\mathbf{U}_{c}'\left(\rho\right).\label{eq:ControlEikonal}
\end{align}
Again we obtain a control function without any reference to the response
function $\mathbf{W}^{\dagger}$. Noteworthy, in order to control
a wave pattern with the proposed method, we solely need to know the
velocity $c$, the invertible coupling matrix $\mathcal{G}$, the
one-dimensional wave profile $\mathbf{U}_{c}$, and the matrix of
diffusion coefficients $\mathcal{D}$.\\
Numerical simulations are typically performed in Cartesian coordinates
$\mathbf{r}=\left(x,y\right)^{T}$. To evaluate the control function
Eq. \eqref{eq:ControlEikonal}, we need to express $\left(\rho,s\right)^{T}$
in terms of the Cartesian coordinates $\left(x,y\right)^{T}$. Thus
we have to invert the coordinate transform Eq. \eqref{eq:CoordinateTransform}
for every time step. In general, this can only be done numerically,
with the Newton-Raphson root finding method as a possible algorithm.\\
Below, we present a specific example. We choose a parametrization
of the curve in polar coordinates
\begin{eqnarray}
\boldsymbol{\gamma}\left(s,t\right) & = & R\left(s,t\right)\left(\begin{array}{c}
\cos\left(s\right)\\
\sin\left(s\right)
\end{array}\right),
\end{eqnarray}
whereby the coordinate $s$ is restricted to $0\leq s<2\pi$. The
normal vector $\mathbf{n}\left(s,t\right)$ and normal velocity $c_{n}$
of the curve are given by
\begin{eqnarray}
\mathbf{n}\left(s,t\right) & = & \frac{\left(\begin{array}{c}
\cos\left(s\right)R'\left(s,t\right)-\sin\left(s\right)R\left(s,t\right)\\
\cos\left(s\right)R\left(s,t\right)+\sin\left(s\right)R'\left(s,t\right)
\end{array}\right)}{\sqrt{R\left(s,t\right)^{2}+\left(R'\left(s,t\right)\right)^{2}}},
\end{eqnarray}
and
\begin{align}
c_{n}\left(s,t\right) & =\mathbf{n}\left(s,t\right)\cdot\dot{\boldsymbol{\gamma}}\left(s,t\right)\nonumber \\
 & =\frac{R\left(s,t\right)\dot{R}\left(s,t\right)}{\sqrt{R\left(s,t\right)^{2}+\left(R'\left(s,t\right)\right)^{2}}},
\end{align}
respectively. For the curvature $\kappa$ follows
\begin{align}
\kappa\left(s,t\right) & =\frac{R\left(s,t\right)^{2}+2\left(R'\left(s,t\right)\right)^{2}-R\left(s,t\right)R''\left(s,t\right)}{\left(R\left(s,t\right)^{2}+\left(R'\left(s,t\right)\right)^{2}\right)^{3/2}}.
\end{align}
To find the new coordinates $\left(\rho,s\right)^{T}$ in terms of
the Cartesian coordinates $\mathbf{r}=\left(x,y\right)^{T}$, we proceed
as follows. At the curve $\boldsymbol{\chi}\left(0,s,t\right)=\boldsymbol{\gamma}\left(s,t\right)$,
the coordinate transform Eq. \eqref{eq:CoordinateTransform} can be
inverted and the coordinate $s$ is given by 
\begin{eqnarray}
s & = & \arctan\left(x,y\right).
\end{eqnarray}
The two-argument function $\arctan\left(x,y\right)$ denotes the arctangent
of $y/x$ within the range $(\lyxmathsym{\textminus}\pi,\pi]$ instead
of $\left(-\pi/2,\pi/2\right)$ as given for the usual $\arctan\left(y/x\right)$.
Close to the curve $\boldsymbol{\chi}\left(0,s,t\right)=\boldsymbol{\gamma}\left(s,t\right)$,
we can expand $\boldsymbol{\chi}\left(\rho,s,t\right)$ around $\left(\rho_{0},s_{0}\right)^{T}$
with $\rho_{0}=0$
\begin{eqnarray}
\boldsymbol{\chi}\left(\rho,s,t\right) & \approx & \boldsymbol{\chi}\left(\rho_{0},s_{0},t\right)+J\left(\rho_{0},s_{0}\right)\cdot\left(\begin{array}{c}
\rho-\rho_{0}\\
s-s_{0}
\end{array}\right)\nonumber \\
 & = & \boldsymbol{\gamma}\left(s_{0},t\right)+J\left(0,s_{0}\right)\cdot\left(\begin{array}{c}
\rho\\
s-s_{0}
\end{array}\right),
\end{eqnarray}
where $J\left(\rho,s\right)$ denotes the Jacobian of the coordinate
transformation Eq. \eqref{eq:CoordinateTransform}. Solving this linear
equation, we find
\begin{eqnarray}
\left(\begin{array}{c}
\rho\\
s
\end{array}\right) & = & \left(\begin{array}{c}
0\\
s_{0}
\end{array}\right)+J^{-1}\left(0,s_{0}\right)\cdot\left(\mathbf{r}-\boldsymbol{\gamma}\left(s_{0},t\right)\right)
\end{eqnarray}
with $s_{0}$ given by $s_{0}=\arctan\left(x,y\right)$. This approximation
is valid close to the curve $\boldsymbol{\gamma}$ but deteriorates
further away from it. Thus, for better accuracy, we apply the Newton-Raphson
iteration
\begin{align}
\left(\begin{array}{c}
\rho_{k+1}\\
s_{k+1}
\end{array}\right) & =\left(\begin{array}{c}
\rho_{k}\\
s_{k}
\end{array}\right)+J^{-1}\left(\rho_{k},s_{k}\right)\cdot\left(\mathbf{r}-\boldsymbol{\gamma}\left(s_{k},t\right)\right),\label{eq:NewtonRaphson}
\end{align}
with the initial values 
\begin{eqnarray}
\rho_{0} & =0,\qquad & s_{0}=\arctan\left(x,y\right).
\end{eqnarray}
In numerical simulations we perform typically 100 iterations of Eq.
\eqref{eq:NewtonRaphson} to achieve a sufficiently accurate result.\\
In the following, we apply the shape control Eq. \eqref{eq:ControlEikonal}
to the FitzHugh-Nagumo (FHN) model \cite{fitzhugh1961impulses,nagumo1962active}
\begin{eqnarray}
\partial_{t}u & = & D_{u}\triangle u+3u-u^{3}-v+\epsilon\left(\mathcal{G}_{11}f_{u}+\mathcal{G}_{12}f_{v}\right),\label{eq:FitzHughNagumo}\\
\partial_{t}v & = & D_{v}\triangle v+\tilde{\epsilon}\left(u-\delta\right)-\tilde{\epsilon}\gamma v+\epsilon\left(\mathcal{G}_{21}f_{u}+\mathcal{G}_{22}f_{v}\right).\nonumber 
\end{eqnarray}
In the absence of control, equations \eqref{eq:FitzHughNagumo} posses
a well known stable traveling pulse solution in one spatial dimension.
No exact analytical expression is known for this solution. Therefore,
we use a numerically determined, linearly interpolated one-dimensional
pulse profile to evaluate $\mathbf{U}_{c}'\left(\rho\right)$ in the
control signal Eq. \eqref{eq:ControlEikonal}. $\mathcal{G}_{ij}$
denotes the components of the coupling matrix $\mathcal{G}$ which
we set to $\mathcal{G}=\mathbf{1}$ for simplicity. We initialize
the controlled pattern sufficiently far away from any domain boundary
so that the Neumann boundary conditions Eq. \eqref{eq:NeumannBoundary}
are approximately satisfied. The parameters for the FHN model are
\begin{align}
\tilde{\epsilon}=0.33,\, & \delta=-1.3,\,\gamma=0,\, D_{u}=1.0,\, D_{v}=0.3.
\end{align}
\begin{figure}
\begin{center}\includegraphics{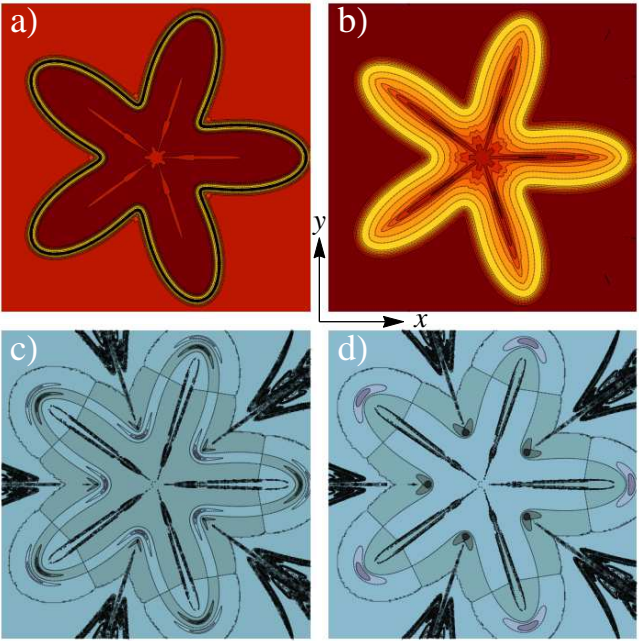}\end{center}

\caption{\label{fig:EikonalShapeControl}(Color online) Shaping a circular
FitzHugh-Nagumo pulse to a flower with five petals. a) activator $u$,
b) inhibitor $v$, c) activator control $f_{u}$, d) inhibitor control
$f_{v}$. The protocol curve $\boldsymbol{\gamma}\left(s,t\right)$
outlining the maximum activator value is shown as the black line (top
left). Small numerical artifacts resulting from inverting the coordinate
transform $\boldsymbol{\chi}$ from Cartesian $\left(x,y\right)$
to local coordinates $\left(\rho,s\right)$ can be seen as a dark
sahdow in the control functions. Domain size is $L_{x}=L_{y}=150$.
See Supplemental Material \cite{supplement} for a movie.}
\end{figure}
Without control, an initially circular wave pattern propagates radially
outwards with a time dependent radius $r\left(t\right)=1/\kappa\left(t\right)$
given by the solution of the unperturbed linear eikonal equation \eqref{eq:LinearEikonalEquation},
\begin{eqnarray}
\dot{r}\left(t\right) & = & c-\frac{\alpha}{r\left(t\right)},
\end{eqnarray}
as
\begin{align}
r\left(t\right) & =\frac{\alpha}{c}w\left(\left(\frac{cR_{0}}{\alpha}-1\right)\exp\left(\frac{c}{\alpha}\left(c\left(t-t_{0}\right)+R_{0}\right)-1\right)\right)\nonumber \\
 & +\frac{\alpha}{c}.
\end{align}
Here, $R_{0}$ is the initial radius and $w\left(z\right)$ denotes
the Lambert $w$ function. We design a control which impedes outward
propagation and deforms the circular pattern into a flower-like shape
with 5 petals. For that purpose we set
\begin{eqnarray}
R\left(s,t\right) & = & R_{0}+\frac{At}{T}\cos\left(ms\right)
\end{eqnarray}
with
\begin{eqnarray}
R_{0} & =50, & \qquad A=40,\qquad m=5,\qquad T=5.
\end{eqnarray}
Fig. \ref{fig:EikonalShapeControl} shows a snapshot of the time evolution
under this control, see the Supplemental Material \cite{supplement}
for a movie. The control deforms the activator, Fig. \ref{fig:EikonalShapeControl}a),
and the inhibitor shape, Fig. \ref{fig:EikonalShapeControl}b), into
a flower like pattern. The position of the traveling wave, given by
the maximum activator value, follows the prescribed curve $\boldsymbol{\gamma}$
(black solid line in Fig. \ref{fig:EikonalShapeControl}a)) closely.
Fig. \ref{fig:EikonalShapeControl}c) and Fig. \ref{fig:EikonalShapeControl}d)
display the corresponding control functions $f_{u}$ and $f_{v}$,
respectively. The control signals attain its largest amplitudes at
the points of largest curvature $\kappa$ along the curve $\boldsymbol{\gamma}$.
Small numerical noise is visible as a dark shadow and results from
the imperfect inversion of the coordinate transform via the Newton-Raphson
algorithm Eq. \eqref{eq:NewtonRaphson}.

\section{\label{sec:PositionControlInFiniteDomains}Position control in finite
domains}

\begin{figure}
\begin{center}\includegraphics{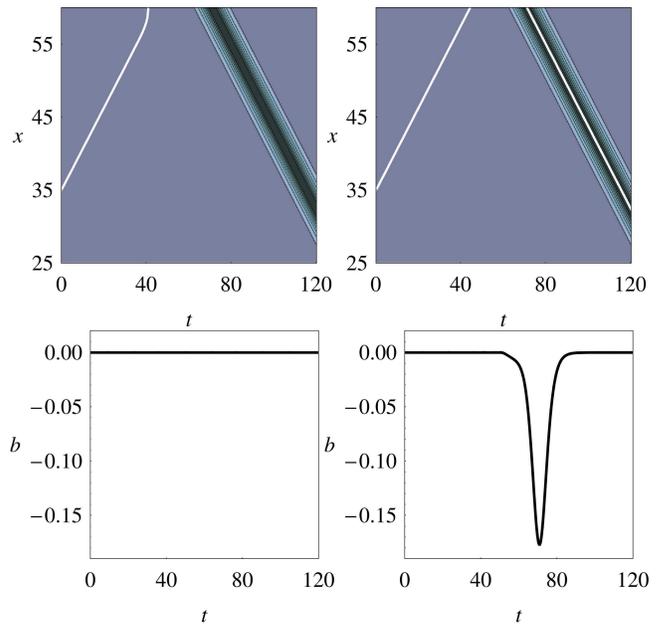}\end{center}\caption{\label{fig:BoundaryControl}(Color online) Additionally to the bulk
control $\mathbf{f}\left(x,t\right)$ acting inside the domain, a
boundary control $b\left(t\right)$ is necessary to move a wave from
inside the domain to the outside and back again. Top: Space-time plot
of the bulk control $\mathbf{f}\left(x,t\right)$ acting inside the
domain. Light (dark) corresponds to a high (low) value of $\mathbf{f}$.
The white line denotes the contour of the Schlögl front with $u\left(x,t\right)=1/2$.
Bottom: Boundary control over time applied at the upper domain boundary
at $x=60$. Left: Under bulk control alone, a front leaving the domain
cannot be returned. Right: With additional boundary control the front
is successfully forced back into the domain. Parameter of the Schlögl
model is $a=0.1$. See Supplemental Material \cite{supplement} for
two movies.}
\end{figure}

Strictly speaking, traveling wave solutions with stationary profile
$\mathbf{U}_{c}$ in a comoving frame can only be defined in an infinite
domain. Any terms in the RD system, including boundary conditions,
which break the translational invariance of the system, destroy the
existence of a traveling wave solution $\mathbf{U}_{c}$. However,
numerical computations of RD systems on finite domains $\Omega$ often
assume homogeneous Neumann boundary conditions 
\begin{eqnarray}
\hat{\mathbf{N}}\left(\mathbf{r}\right)\cdot\nabla\mathbf{u}\left(\mathbf{r},t\right) & = & 0,\,\mathbf{r}\in\Gamma=\partial\Omega,\label{eq:NeumannBoundaries}
\end{eqnarray}
where $\Gamma$ denotes the domain boundary, $\hat{\mathbf{N}}$ is
the unit vector normal to the domain boundary, and the gradient $\nabla$
acts component-wise on the vector $\mathbf{u}$. Physically, Neumann
boundary conditions describe a vanishing flux of components $\mathbf{u}$
across the boundary. Traveling wave solutions to RD systems are typically
localized in the sense that the derivatives of any order $n\geq1$
of the wave profile $\mathbf{U}_{c}\left(\xi\right)$ with respect
to the traveling wave coordinate $\xi$ decay to zero,
\begin{eqnarray}
\lim_{\xi\rightarrow\pm\infty}\partial_{\xi}^{n}\mathbf{U}_{c}\left(\xi\right) & = & 0.\label{eq:LocalizationTravelingWave}
\end{eqnarray}
Because of the localization Eq. \eqref{eq:LocalizationTravelingWave},
the Neumann boundary conditions are approximately satisfied if the
position of traveling waves is sufficiently far away from the boundaries
such that the wave pattern is unaffected by the Neumann boundary.
Sufficiently close to a Neumann boundary, the wave interacts with
the boundary, and our proposed bulk control functions $\mathbf{f}$,
acting inside the domain $\Omega$, might fail because it was derived
under the assumption of an unbounded domain. In this section we show
that this is indeed the case, and demonstrate that the application
of an additional boundary control successfully restores position control
of traveling waves.\\
Assuming that the in- or outflux $\mathbf{b}$ of components across
the boundary can be controlled, we introduce an inhomogeneous Neumann
boundary condition with a boundary control term $\mathbf{b}\left(\mathbf{r},t\right)$
as inhomogeneity,
\begin{eqnarray}
\hat{\mathbf{N}}\left(\mathbf{r}\right)\cdot\nabla\mathbf{u}\left(\mathbf{r},t\right) & = & \mathbf{b}\left(\mathbf{r},t\right),\,\mathbf{r}\in\Gamma.
\end{eqnarray}
Enforcing a desired distribution inside the domain solely by the boundary
control $\mathbf{b}$ is important for applications and can be achieved
by optimal control, see e.g. \cite{PhysRevLett.91.208301} for an
example and \cite{troltzsch2010optimal,theissen2006optimale} for
the general approach. However, here we assume that the additionally
to the bulk control signal $\mathbf{f}\left(x,t\right)$ inside the
domain $\Omega$ a control $\mathbf{b}$ acts on the domain boundary.\\
For simplicity we consider a one-dimensional RD system. A generalization
to higher spatial dimensions is straightforward. Due to the finite
spatial domain the traveling wave profile $\mathbf{U}_{c}\left(x-ct\right)$
ceases to be a solution of the unperturbed RD system Eq. \eqref{eq:ControlledRDS}
with $\epsilon=0$ . However, it becomes again an exact solution if
we supplement the RD system with the inhomogeneous Neumann boundary
condition 
\begin{eqnarray}
\partial_{x}\mathbf{u}\left(x,t\right) & = & \mathbf{U}_{c}'\left(x-ct\right),\, x\in\Gamma.
\end{eqnarray}
These considerations lead to a generalization of position control
for traveling waves in finite domains. The controlled system now reads,
with $\epsilon=1$ as before,
\begin{align}
\partial_{t}\mathbf{u}\left(x,t\right) & =\mathcal{D}\partial_{x}^{2}\mathbf{u}+\mathbf{R}\left(\mathbf{u}\right)+\epsilon\mathbf{f}\left(x,t\right),\, x\in\Omega\subset\mathbb{R}\\
\partial_{x}\mathbf{u}\left(x,t\right) & =\mathbf{U}_{c}'\left(x-ct\right)+\epsilon\mathbf{b}\left(x,t\right),\, x\in\Gamma.
\end{align}
The bulk control follows from the one-dimensional version of Eq. \eqref{eq:TwoDimControl2}
and Eq. \eqref{eq:ControlEikonal}, see also \cite{loeber2013}, 
\begin{eqnarray}
\mathbf{f}\left(x,t\right) & = & \left(c-\dot{\phi}\left(t\right)\right)\mathbf{U}_{c}'\left(x-\phi\left(t\right)\right),\label{eq:BulkControl}
\end{eqnarray}
while the boundary control is given by 
\begin{eqnarray}
\mathbf{b}\left(x,t\right) & = & \mathbf{U}_{c}'\left(x-\phi\left(t\right)\right)-\mathbf{U}_{c}'\left(x-ct\right),\, x\in\Gamma.\label{eq:BoundaryControl}
\end{eqnarray}
This yields the inhomogeneous Neumann boundary condition intuitively
expected for a traveling wave shifted according to the protocol: 
\begin{eqnarray}
\partial_{x}\mathbf{u}\left(x,t\right) & = & \mathbf{U}_{c}'\left(x-\phi\left(t\right)\right),\, x\in\Gamma.
\end{eqnarray}
Fig. \eqref{fig:BoundaryControl} proves the need for boundary control
in the case of the one-dimensional Schlögl model, Eq. \eqref{eq:SchloeglModel}
in one spatial dimension. The task is to move the front across the
boundary according to the protocol
\begin{align}
\phi\left(t\right) & =\phi_{0}+\Theta\left(t_{1}-t\right)\frac{c\left(t_{1}-t_{0}\right)}{\pi}\sin\left(\pi\frac{t-t_{0}}{t_{1}-t_{0}}\right)\nonumber \\
 & -c\left(t-t_{1}\right)\Theta\left(t-t_{1}\right),
\end{align}
where $\Theta$ denotes the Heaviside step function. Bulk as well
as boundary control are switched on at $t=t_{0}$ upon which the unperturbed
front moves with velocity $c$. The protocol smoothly reverses the
propagation direction at a time instant when the front is located
outside the domain. After the time $t=t_{1}$, the front moves backwards
with velocity $-c$. For numerical simulations we set $\phi_{0}=63.28,\, t_{1}-t_{0}=15,$
and $L=60$ for the domain size. The white line in the top panel of
Fig. \ref{fig:BoundaryControl} shows a space-time plot of the position
of the traveling wave determined by $u\left(x,t\right)=1/2$. In Fig.
\ref{fig:BoundaryControl} left, only the bulk control Eq. \eqref{eq:BulkControl}
is applied. A front which has moved beyond the domain boundary cannot
be forced back into the domain. On the contrary, under both bulk and
boundary control the front can successfully be returned, see Fig.
\ref{fig:BoundaryControl} right. The top panels show the spatio-temporal
bulk control $\mathbf{f}\left(x,t\right)$ (gray) and a contour of
the Schlögl front (white), while the bottom panels show the boundary
control $b$ over time, i.e., Eq. \eqref{eq:BoundaryControl} evaluated
at the domain boundary $x=L$.

\section{\label{sec:Conclusions}Conclusions}

We have presented a method to control the shape of two-dimensional
traveling waves in RD systems. The desired wave shape is prescribed
by a time-dependent curve $\boldsymbol{\gamma}$. An inverse problem
for the determination of the control signal $\mathbf{f}$ that forces
the wave to adopt the shape outlined by $\boldsymbol{\gamma}$ is
formulated in terms of well-known evolution equations for wave patterns,
namely the linear eikonal equation and the nonlinear phase diffusion
equation. The feasibility of the proposed approach is demonstrated
by numerical simulations of simple but representative examples of
controlled RD systems.\\
Two variants of this control method with different areas of application
are presented. The first one is suited to shape wave patterns close
to plane waves. In this case the control signal, Eq. \eqref{eq:TwoDimControl2},
is given as the solution of an integral equation based on the perturbed
nonlinear phase diffusion equation \eqref{eq:PerturbedPhaseDiffusionEquation}.
The corresponding desired concentration fields enforced by the control
are 
\begin{eqnarray}
\mathbf{u}_{d}\left(\mathbf{r},t\right) & = & \mathbf{U}_{c}\left(x-\phi\left(y,t\right)\right),\label{eq:DesiredDistribution1}
\end{eqnarray}
where $\mathbf{U}_{c}$ is the uncontrolled one-dimensional wave profile
and $\phi\left(y,t\right)$ denotes the $x$-component of the curve
$\boldsymbol{\gamma}\left(y,t\right)=\left(\phi\left(y,t\right),y\right)^{T}$outlining
the wave shape. Because $\boldsymbol{\gamma}$ is parametrized in
Cartesian coordinates, the control signal $\mathbf{f}$, Eq. \eqref{eq:TwoDimControl2},
can be readily evaluated.\\
The second control method allows to enforce a wave shape described
by a curve $\boldsymbol{\gamma}$ which is not necessarily close to
a plane wave but has a sufficiently small curvature such that the
linear eikonal equation \eqref{eq:LinearEikonalEquation} is valid.
The control signal Eq. \eqref{eq:ControlEikonal} is obtained from
the perturbed eikonal equation \eqref{eq:PerturbedLinearEikonalEquation}
which is interpreted as an integral equation for the unknown control
$\mathbf{f}$ enforcing a wave shape $\boldsymbol{\gamma}$ specified
in advance. The corresponding desired concentration fields are 
\begin{eqnarray}
\mathbf{u}_{d}\left(\mathbf{r},t\right) & = & \mathbf{U}_{c}\left(\rho\right),\label{eq:DesiredDistribution2}
\end{eqnarray}
with $\rho=\rho\left(\mathbf{r},t\right)$ given by the inverse of
the coordinate transformation Eq. \eqref{eq:CoordinateTransform}.
To evaluate the control signal $\mathbf{f}\left(\boldsymbol{\chi}\left(\rho,s,t\right),t\right)$,
this inverse coordinate transformation must be computed for every
time step. This renders the second control method computationally
much more expensive than the first.\\
In principle one could quantify the performance of a certain control
signal by evaluating the squared difference between the desired concentration
fields $\mathbf{u}_{d}$, Eq. \eqref{eq:DesiredDistribution1} or
Eq. \eqref{eq:DesiredDistribution2}, and the actual numerical result
$\mathbf{u}$ of the controlled RD system and integrate the result
over the spatiotemporal computational domain $\left[0,L_{x}\right]\times\left[0,L_{y}\right]\subset\mathbb{R}^{2}$,
\begin{align}
S & =\frac{1}{\left(t_{1}-t_{0}\right)L_{x}L_{y}}\intop_{t_{0}}^{t_{1}}dt\intop_{0}^{L_{x}}dx\intop_{0}^{L_{y}}dy\left(\mathbf{u}\left(\mathbf{r},t\right)-\mathbf{u}_{d}\left(\mathbf{r},t\right)\right)^{2}.\label{eq:SquaredDifference}
\end{align}
A constrained functional as in Eq. \eqref{eq:SquaredDifference} would
also be the starting point for an optimal control algorithm, i.e.,
an iterative algorithm which aims to minimize $S$ \cite{buchholz2013on,troltzsch2010optimal,jorge1999numerical}.\\
We emphasize that the control signal $\mathbf{f}$ is always expressed
solely in terms of the derivative $\mathbf{U}_{c}'\left(\xi\right)$
of the uncontrolled one-dimensional traveling wave profile, its propagation
velocity $c$, the diagonal matrix of diffusion coefficients $\mathcal{D}$
and the invertible coupling matrix $\mathcal{G}$. We are able to
eliminate any reference to the response function $\mathbf{W}^{\dagger}\left(x\right)$
and consequently, to apply our method it is not necessary to know
the reaction kinetics $\mathbf{R}$. This makes the approach useful
for applications where the underlying reaction kinetics is not at
all or only approximately known, while in the absence of control,
the traveling wave profile $\mathbf{U}_{c}$ of all state variables
and its propagation velocity $c$ can be measured with sufficient
accuracy. In case that only an incomplete number of (combinations
of) state variables can be measured, observer techniques from nonlinear
control theory \cite{krener1983linearization,khalil2002nonlinear}
or other state estimation and fitting procedures can be applied to
reconstruct the complete state. However, this usually requires a more
detailed knowledge of the reaction kinetics. In contrast to feedback
control, which necessitates the continuous estimation of the complete
state of the actual system to be controlled, for our approach it is
sufficient to obtain the wave profile of all state variables from
an identical copy of the uncontrolled system. Therefore, a much larger
variety of measurement techniques, as e.g. destructive measurements,
can be deployed, and the notion of nonlinear state observability \cite{hermann1977automatic}
developed for feedback control does not apply in our case. Once the
wave profile $\mathbf{U}_{c}$ is obtained with sufficient accuracy,
the application of our control scheme merely requires the knowledge
of the initial shape of the wave pattern to be controlled, which can
be obtained from the measurement of a single state component.\\
In the derivation of the perturbed eikonal and phase diffusion equations,
Eq. \eqref{eq:PerturbedLinearEikonalEquation} and Eq. \eqref{eq:PerturbedPhaseDiffusionEquation},
respectively, we assumed a small amplitude of the control signal $\mathbf{f}$.
Note that the control amplitude ${c-c_{n}-\alpha\kappa}$ in Eq. \eqref{eq:ControlEikonal0}
is zero if and only if the time evolution of the curve $\boldsymbol{\ensuremath{\gamma}}$
is governed by the unperturbed linear eikonal equation \eqref{eq:LinearEikonalEquation}.
In other words, trying to enforce a wave shape which is a solution
of the unperturbed RD equations leads to a vanishing control signal
$\mathbf{f}$.\\
In the examples discussed so far the coupling matrix is restricted
to the simplest possible case ${\mathcal{G}=\mathbf{1}}$. However,
our method can be readily generalized to more complex but invertible
coupling matrices. For multi-component systems, invertibility of $\mathcal{G}$
implies that the number $m$ of independent control signals ${\mathbf{f}\left(\mathbf{r},t\right)=\left(f_{1}\left(\mathbf{r},t\right),\dots,f_{m}\left(\mathbf{r},t\right)\right)}$
is equal to the number $n$ of components ${\mathbf{u}\left(\mathbf{r},t\right)=\left(u_{1}\left(\mathbf{r},t\right),\dots,u_{n}\left(\mathbf{r},t\right)\right)^{T}}$
of the RD system. Using the control solution ${\mathbf{f}\left(\mathbf{r},t\right)=\left(f_{u}\left(\mathbf{r},t\right),f_{v}\left(\mathbf{r},t\right)\right)^{T}}$,
Eq. \eqref{eq:ControlEikonal}, for the controlled Fitz-Hugh Nagumo
model, Eq. \eqref{eq:FitzHughNagumo}, with ${\mathcal{G}=\mathbf{1}}$,
we demonstrate how the method can be generalized if the coupling matrix
$\mathcal{G}$ is not invertible and the control $\tilde{f}_{u}$
is acting solely on the activator variable, 
\begin{eqnarray}
\partial_{t}u & = & D_{u}\triangle u+3u-u^{3}-v+\epsilon\tilde{f}_{u},\label{eq:SingularFitzHughNagumo}\\
\partial_{t}v & = & D_{v}\triangle v+\tilde{\epsilon}\left(u-\delta\right)-\tilde{\epsilon}\gamma v.\nonumber 
\end{eqnarray}
The FitzHugh-Nagumo model Eq. \eqref{eq:FitzHughNagumo} with ${\mathcal{G}=\mathbf{1}}$
can be written as a single nonlinear integrodifferential equation
for the activator $u$,
\begin{eqnarray}
\partial_{t}u & = & D_{u}\triangle u+3u-u^{3}-\mathcal{K}\left(\tilde{\epsilon}\left(u-\delta\right)+\epsilon f_{v}\right)\nonumber \\
 &  & -\mathcal{K}_{0}v_{0}+\epsilon f_{u}.
\end{eqnarray}
Here, $\mathcal{K}$ and $\mathcal{K}_{0}$ are integral operators,
involving Green's function, of the inhomogeneous linear PDE for the
inhibitor $v$ with initial condition $v\left(\mathbf{r},t_{0}\right)=v_{0}\left(\mathbf{r}\right)$,
\begin{align}
\partial_{t}v-D_{v}\triangle v+\tilde{\epsilon}\gamma v & =\tilde{\epsilon}\left(u-\delta\right)+\epsilon f_{v}.
\end{align}
Equation \eqref{eq:SingularFitzHughNagumo} can be written as single
nonlinear integrodifferential equation as well. Comparing the control
terms of both integrodifferential equations, we obtain a solution
for the control $\tilde{f}_{u}$ as 
\begin{align}
\tilde{f}_{u}\left(\mathbf{r},t\right) & =-\mathcal{K}f_{v}\left(\mathbf{r},t\right)+f_{u}\left(\mathbf{r},t\right).
\end{align}
The term $h\left(\mathbf{r},t\right)=\mathcal{K}f_{v}\left(\mathbf{r},t\right)$
can be computed as the solution to the inhomogeneous PDE
\begin{align}
\partial_{t}h-D_{v}\triangle h+\tilde{\epsilon}\gamma h & =f_{v}\left(\mathbf{r},t\right)
\end{align}
with initial condition $h\left(\mathbf{r},t_{0}\right)=0$. This computation
requires more knowlegde of the underlying reaction kinetics when compared
to the case of an invertible coupling matrices, namely the values
of the parameters $\tilde{\epsilon}$ and $\gamma$ must be known.
See \cite{loeber2013} and the accompanying supplement for one-dimensional
numerical simulations of the controlled RD system with singular coupling
matrix, Eq. \eqref{eq:SingularFitzHughNagumo}, and a comparison with
optimal control as well as a generalization to Hodgkin-Huxley type
and other RD models.\\
An important aspect of the proposed control method is stability. Stability
means that the controlled reaction-diffusion system \eqref{eq:ControlledRDS}
must yield a solution which is sufficiently close to the desired distribution,
Eq. \eqref{eq:DesiredDistribution1} or Eq. \eqref{eq:DesiredDistribution2}.
For position control of traveling waves in one spatial dimension,
the controlled reaction-diffusion system \eqref{eq:ControlledRDS}
with control function Eq. \eqref{eq:BulkControl}, 
\begin{align}
\partial_{t}\mathbf{u} & =\mathcal{D}\partial_{x}^{2}\mathbf{u}+\mathbf{R}\left(\mathbf{u}\right)+\left(c-\dot{\theta}\left(t\right)\right)\mathbf{U}_{c}'\left(x-\theta\left(t\right)\right)\nonumber \\
 & +\hat{\epsilon}\mathbf{q}\left(\mathbf{u},x,t\right),
\end{align}
must be stable against structural perturbations $\mathbf{q}\left(\mathbf{u},x,t\right)$
of the system equations itself as well as stable against perturbations
$\mathbf{z}_{0}$ and $\Delta X$ of the initial conditions,
\begin{align}
\mathbf{u}\left(x,t_{0}\right) & =\mathbf{U}_{c}\left(x-\theta_{0}-\Delta X\right)+\hat{\epsilon}\mathbf{z}_{0}\left(x\right).
\end{align}
The latter arise if e.g. the initial condition is not exactly the
traveling wave solution $\mathbf{U}_{c}$, or if the control is applied
initially at a position $\theta_{0}=\theta\left(t_{0}\right)$ different
from the actual initial position $\theta_{0}+\Delta X$ of the traveling
wave to be controlled. Structural perturbations $\mathbf{q}$ could
be spatial heterogeneities of the reaction-diffusion medium, for example,
or errors in measurements of the traveling wave profile $\mathbf{U}_{c}$
which cause an incorrect and noisy control function. For a quantitative
analysis, the parameter $\hat{\epsilon}\ll1$ is assumed to be small
such that the system can be linearized with an appropriate ansatz.
Note that $\hat{\epsilon}$ is not necessarily of the same magnitude
as the original small parameter $\epsilon$ in Eq. \eqref{eq:ControlledRDS},
which we set to $\epsilon=1$ here. The solution $\mathbf{v}$ to
the linear equations can be written as a superposition of eigenfunctions
$\mathbf{v}_{i}$ of the linear operator $\mathcal{L}$, Eq. \eqref{eq:LinearOperatorL},
weighted by factors $\sim\exp\left(-\lambda_{i}t\right)$ \cite{loeber2013},
where $\lambda_{i}$ is the $i$-th eigenvalue of $\mathcal{L}$.
If the traveling wave solution $\mathbf{U}_{c}$ is stable and exhibits
a spectral gap, then all eigenvalues for $i>0$ have a real part $\Re\left(\lambda_{i}\right)<0$.
Therefore, apart from the Goldstone mode $\mathbf{v}_{0}=\mathbf{W}=\mathbf{U}_{c}'$
corresponding to the zero eigenvalue $\lambda_{0}=0$, any perturbation
$\hat{\epsilon}\mathbf{z}_{0}$ of the initial conditions decays to
zero for large times. The larger is the spectral gap, the faster decays
the perturbation. Similarly, any structural perturbation $\hat{\epsilon}\mathbf{q}\left(\mathbf{u},x,t\right)$
decays to a term with magnitude proportional to $\hat{\epsilon}$,
leading merely to a small time-dependent deformation of the traveling
wave profile $\mathbf{U}_{c}$ as long as $\hat{\epsilon}$ is sufficiently
small. Because the Goldstone mode with eigenvalue $\lambda_{0}=0$
does not decay in time, it requires special treatment. In \cite{loeber2013Stability},
we derived and analyzed an EOM very similar to Eq. \eqref{eq:1DimEquationOfMotion}
which takes into account the dynamics of the Goldstone mode. This
analysis established the stability of position control against perturbations
$\Delta X$ of the initial position in the absence of structural perturbations.
In accordance with numerical simulations, we found that an interval
of perturbations $\Delta X$ exists for which position control is
stable. The size of this interval is a measure for the stability against
perturbations of the initial position, and, in general, depends on
the reaction kinetics in a complicated way. The interval vanishes
for stationary traveling waves with reflection symmetry. As for the
effect of structural perturbations $\mathbf{q}\left(\mathbf{u},x,t\right)$,
we expect that as long as the parameter $\hat{\epsilon}$ characterizing
the amplitude of structural perturbations is sufficiently small, this
interval does still exist. In principle, structural perturbations
cannot only shrink the interval but also enlarge it, thereby exerting
a destabilizing or stabilizing influence on position control, respectively.
Going from one to two spatial dimensions, new phenomena as e.g. transversal
instabilities arise \cite{kuramoto1980instability,kuramoto2003chemical}
which can destabilize an uncontrolled plane wave. However, we expect
that our qualitative discussion of stability also applies in this
case provided a plane wave is stable and exhibits a spectral gap.
A rigorous quantitative justification of the arguments above is quite
involved and set aside for future publications.\\
Generalizations of the linear eikonal equation and the nonlinear
phase diffusion equation to three spatial dimensions describe the
evolution of isoconcentration surfaces. We expect that along the lines
of the approach discussed in the present paper it will be possible
to control the shape of three-dimensional wave patterns in RD systems.\\
In view of the growing relevance of RD models in such diverse fields
as (bio-)chemical reactions, population dynamics \cite{turchin2003complex},
the cooperative self-organization of microorganisms \cite{benjacob2000cooperative},
infectious diseases \cite{bailey1987mathematical}, and physiology
\cite{sneyd2008mathematical}, imaginable applications for our proposed
control method abound. We mention the prevention of the spreading
of epidemics \cite{murray1993mathematical}, guiding of chemically
propelled nanomotors interacting with chemical waves \cite{thakur2011dynamics,thakur2011interaction},
the growth of crystals into desired shapes \cite{burton1951thegrowth,cartwright2012crystalgrowth},
and the control of evolving cell cultures as e.g. tumor progression
\cite{ferreira2002reaction,araujo2004ahistory}. Possible experimental
systems to test the feasibility of the proposed control method are
the light-sensitive BZ reaction \cite{krug1990analysis} or a liquid
crystal light valve with optical feedback \cite{residori2005nonlinear,haudin2009driven,haudin2010front},
see also the discussion in \cite{loeber2014control}.

\appendix

\section{\label{sec:DerivationPNPDE}Derivation of the perturbed nonlinear
phase diffusion equation}

We start with the perturbed RD system

\begin{eqnarray}
\partial_{t}\mathbf{u} & = & \mathcal{D}\triangle\mathbf{u}+\mathbf{R}\left(\mathbf{u}\right)+\epsilon\mathbf{f}\left(\mathbf{u},x,y,t\right),
\end{eqnarray}
where $\triangle=\partial_{x}^{2}+\partial_{y}^{2}$ is the Laplacian
and $\mathcal{D}$ is a diagonal matrix of constant diffusion coefficients.
The space domain extends from $-\infty$ to $\infty$ in the $x$-direction.
In the $y$-direction it is either finite with no flux or periodic
boundary conditions, or infinite as well. The goal is to derive an
equation for the shape $\phi\left(y,t\right)$ of the wave over time.
The unperturbed solution is assumed to be a traveling wave $\mathbf{U}_{c}\left(x-ct\right)$
propagating with constant velocity $c$ in the $x$-direction. The
wave profile is stationary in the comoving frame with $\xi=x-ct$,
i.e., $\mathbf{U}_{c}$ obeys 
\begin{eqnarray}
\mathcal{D}\mathbf{U}_{c}''\left(\xi\right)+c\mathbf{U}_{c}'\left(\xi\right)+\mathbf{R}\left(\mathbf{U}_{c}\left(\xi\right)\right) & = & 0.
\end{eqnarray}
We introduce a slow time scale, $T=\epsilon t$, and a stretched spatial
coordinate in the $y$-direction, $Y=\epsilon^{1/2}y$. Bearing in
mind the replacements
\begin{align}
\partial_{t}\rightarrow\partial_{t}+\partial_{t}T\partial_{T} & =\partial_{t}+\epsilon\partial_{T},\label{eq:dT}\\
\partial_{y}\rightarrow\partial_{y}+\partial_{y}Y\partial_{Y} & =\partial_{y}+\epsilon^{1/2}\partial_{Y},\\
\partial_{y}^{2}\rightarrow\left(\partial_{y}+\epsilon^{1/2}\partial_{Y}\right)\left(\partial_{y}+\epsilon^{1/2}\partial_{Y}\right) & =\partial_{y}^{2}+2\epsilon^{1/2}\partial_{y}\partial_{Y}+\epsilon\partial_{Y}^{2},\label{eq:dY}
\end{align}
and the ansatz for the solution
\begin{align}
\mathbf{u}\left(x,y,t,Y,T\right) & =\mathbf{U}_{c}\left(x-ct+p\left(Y,T\right)\right)\nonumber \\
 & +\epsilon\mathbf{v}\left(x-ct+p\left(Y,T\right),y,t,Y,T\right),\label{eq:Ansatz}
\end{align}
we derive a PDE for $p\left(Y,T\right)$. We get
\begin{align}
\partial_{t}\mathbf{u}+\epsilon\partial_{T}\mathbf{u} & =\mathcal{D}\triangle\mathbf{u}+2\mathcal{D}\epsilon^{1/2}\partial_{y}\partial_{Y}\mathbf{u}\nonumber \\
 & +\epsilon\mathcal{D}\partial_{Y}^{2}\mathbf{u}+\mathbf{R}\left(\mathbf{u}\right)+\epsilon\mathbf{f}\left(\mathbf{u},x,y,t\right),\label{eq:PDENewScales}
\end{align}
and for the derivatives in $y$-direction up to $\mathcal{O}\left(\epsilon\right)$
\begin{eqnarray}
\epsilon^{1/2}\partial_{y}\partial_{Y}\mathbf{u} & =\epsilon^{1/2}\partial_{Y}\partial_{y}\mathbf{u}= & 0+\mathcal{O}\left(\epsilon^{3/2}\right).
\end{eqnarray}
Expanding Eq. \eqref{eq:PDENewScales} with the ansatz Eq. \eqref{eq:Ansatz}
up to $\mathcal{O}\left(\epsilon^{3/2}\right)$ yields
\begin{align}
\epsilon\mathbf{U}_{c}'\partial_{T}p+\epsilon\partial_{t}\mathbf{v} & =\mathcal{D}\mathbf{U}_{c}''+c\mathbf{U}_{c}'+\mathbf{R}\left(\mathbf{U}_{c}\right)\\
 & +\epsilon\mathcal{D}\mathbf{U}_{c}''\left(\partial_{Y}p\right)^{2}+\epsilon\mathcal{D}\mathbf{U}_{c}'\partial_{Y}^{2}p+\epsilon\mathcal{D}\triangle\mathbf{v}\nonumber \\
 & +\epsilon c\partial_{\xi}\mathbf{v}+\epsilon\mathbb{D}\mathbf{R}\left(\mathbf{U}_{c}\right)\mathbf{v}+\epsilon\mathbf{f}+\mathcal{O}\left(\epsilon^{3/2}\right).\nonumber 
\end{align}
Here, $\mathbb{D}\mathbf{R}\left(\mathbf{U}_{c}\right)$ denotes the
Jacobian of $\mathbf{R}$ and we have transformed to the comoving
coordinate $\xi=x-ct$. The control reads now
\begin{align}
\mathbf{f} & =\mathbf{f}\left(\mathbf{U}_{c}\left(\xi+p\left(Y,T\right)\right),\xi+ct,y,t\right)+\mathcal{O}\left(\epsilon\right).
\end{align}
Simplifying leads to the following equation in order $\mathcal{O}\left(\epsilon\right)$
\begin{align}
\mathbf{U}_{c}'\partial_{T}p & =\mathcal{D}\mathbf{U}_{c}''\left(\partial_{Y}p\right)^{2}+\mathcal{D}\mathbf{U}_{c}'\partial_{Y}^{2}p-\partial_{t}\mathbf{v}+\mathcal{L}\mathbf{v}+\mathbf{f},
\end{align}
where we have introduced
\begin{align}
\mathcal{L} & =\mathcal{D}\left(\partial_{\xi}^{2}+\partial_{y}^{2}\right)+c\partial_{\xi}+\mathbb{D}\mathbf{R}\left(\mathbf{U}_{c}\left(\xi+p\left(Y,T\right)\right)\right).
\end{align}
The operator $\mathcal{L}$ acts on the function ${\mathbf{v}=\mathbf{v}\left(\xi+p\left(Y,T\right),y,Y,T\right)}$.
Now let us suppose we know the eigenfunction $\mathbf{W}^{\dagger}$
to the eigenvalue zero of the operator $\mathcal{L}^{\dagger}$ adjoint
to $\mathcal{L}$ with respect to the standard inner product in function
space 
\begin{eqnarray}
\left\langle \mathbf{w}\left(\xi\right),\mathbf{v}\left(\xi\right)\right\rangle  & = & \intop_{-\infty}^{\infty}d\xi\mathbf{w}^{T}\left(\xi\right)\mathbf{v}\left(\xi\right),
\end{eqnarray}
where $\mathbf{w}{}^{T}$ denotes the transpose of vector $\mathbf{w}$.
${\mathbf{W}^{\dagger}=\mathbf{W}^{\dagger}\left(\xi+p\left(Y,T\right)\right)}$
is called the adjoint Goldstone mode. In general, and particularly
in higher spatial dimensions, there can be more than one adjoint Goldstone
mode. The second contribution $\mathbf{v}\left(x-ct+p\left(Y,T\right),y,t,Y,T\right)$
of the ansatz Eq.\eqref{eq:Ansatz} involves a sum over all eigenfunctions
of $\mathcal{L}$. However, the first contribution $\mathbf{U}_{c}\left(x-ct+p\left(Y,T\right)\right)$
of Eq.\eqref{eq:Ansatz} includes already the effect of the Goldstone
mode $\mathbf{U}_{c}'$ because of $\mathbf{U}_{c}\left(\xi+p\right)\approx\mathbf{U}_{c}\left(\xi\right)+p\mathbf{U}_{c}'\left(\xi\right)$.
Therefore, to exclude the Goldstone mode from $\mathbf{v}$, we have
\begin{eqnarray}
\left\langle \mathbf{W}^{\dagger},\mathbf{v}\right\rangle  & = & 0,
\end{eqnarray}
from which follows 
\begin{eqnarray}
\left\langle \mathbf{W}^{\dagger},\partial_{t}\mathbf{v}\right\rangle  & = & 0
\end{eqnarray}
because $\mathbf{W}^{\dagger}$ is independent of $t$. So we find
\begin{align}
-\left\langle \mathbf{W}^{\dagger},\mathcal{L}\mathbf{v}\right\rangle  & =-\left\langle \mathbf{W}^{\dagger},\mathbf{U}_{c}'\right\rangle \partial_{T}p+\left\langle \mathbf{W}^{\dagger},\mathcal{D}\mathbf{U}_{c}''\right\rangle \left(\partial_{Y}p\right)^{2}\nonumber \\
 & +\left\langle \mathbf{W}^{\dagger},\mathcal{D}\mathbf{U}_{c}'\right\rangle \partial_{Y}^{2}p+\left\langle \mathbf{W}^{\dagger},\mathbf{f}\right\rangle .\label{eq:A1E17}
\end{align}
Eq. \eqref{eq:A1E17} is also called a solvability condition or Fredholm
alternative. Because of $\left\langle \mathbf{W}^{\dagger},\mathcal{L}\mathbf{v}\right\rangle =\left\langle \mathcal{L}^{\dagger}\mathbf{W}^{\dagger},\mathbf{v}\right\rangle =0$,
the right hand side must be zero, and we obtain the desired PDE for
$p$ as follows: 
\begin{align}
0 & =-\left\langle \mathbf{W}^{\dagger},\mathbf{U}_{c}'\right\rangle \partial_{T}p+\left\langle \mathbf{W}^{\dagger},\mathcal{D}\mathbf{U}_{c}''\right\rangle \left(\partial_{Y}p\right)^{2}\nonumber \\
 & +\left\langle \mathbf{W}^{\dagger},\mathcal{D}\mathbf{U}_{c}'\right\rangle \partial_{Y}^{2}p+\left\langle \mathbf{W}^{\dagger},\mathbf{f}\right\rangle .
\end{align}
Note that with our ansatz for ${\mathbf{v}=\mathbf{v}\left(\xi+p\left(T,Y\right),t,y,T,Y\right)}$,
we have 
\begin{eqnarray}
\left\langle \mathbf{W}^{\dagger},\mathbf{v}\right\rangle  & = & \intop_{-\infty}^{\infty}d\xi\mathbf{W}^{\dagger T}\left(\xi+p\right)\mathbf{v}\left(\xi+p,y,t,Y,T\right)\nonumber \\
 & = & \intop_{-\infty}^{\infty}d\xi\mathbf{W}^{\dagger T}\left(\xi\right)\mathbf{v}\left(\xi,y,t,Y,T\right)=0,
\end{eqnarray}
i.e., these inner products do not depend on $p$ any more. An exception
is the inner product involving the perturbation $\mathbf{f}$, which
we transform back to an integral over original space $\tilde{x}$
as follows,
\begin{align}
\left\langle \mathbf{W}^{\dagger},\mathbf{f}\right\rangle  & =\intop_{-\infty}^{\infty}d\xi\mathbf{W}^{\dagger T}\left(\xi+p\right)\mathbf{f}\left(\mathbf{U}_{c}\left(\xi+p\right),\xi+ct,y,t\right)\nonumber \\
 & =\intop_{-\infty}^{\infty}d\tilde{x}\mathbf{W}^{\dagger T}\left(\tilde{x}\right)\mathbf{f}\left(\mathbf{U}_{c}\left(\tilde{x}\right),\tilde{x}+ct-p,y,t\right).
\end{align}
In general, 
\begin{eqnarray}
\frac{\left\langle \mathbf{W}^{\dagger},\mathcal{D}\mathbf{U}_{c}'\right\rangle }{\left\langle \mathbf{W}^{\dagger},\mathbf{U}_{c}'\right\rangle } & \neq & \mathcal{D},
\end{eqnarray}
since $\mathcal{D}$ is a matrix of diffusion coefficients. Only if
the diffusion coefficient is the same for all components, $\mathcal{D}=D\mathbf{1}$,
we have 
\begin{eqnarray}
\frac{\left\langle \mathbf{W}^{\dagger},\mathcal{D}\mathbf{U}_{c}'\right\rangle }{\left\langle \mathbf{W}^{\dagger},\mathbf{U}_{c}'\right\rangle }=\frac{\left\langle \mathbf{W}^{\dagger},D\mathbf{U}_{c}'\right\rangle }{\left\langle \mathbf{W}^{\dagger},\mathbf{U}_{c}'\right\rangle } & = & D.
\end{eqnarray}
We introduce a new function
\begin{eqnarray}
\phi\left(y,t\right) & = & ct-p\left(Y,T\right)
\end{eqnarray}
with (see Eqs. \eqref{eq:dT}, \eqref{eq:dY})
\begin{eqnarray}
\partial_{t}\phi\left(y,t\right) & = & c-\epsilon\partial_{T}p\left(Y,T\right),\\
\partial_{y}\phi\left(y,t\right) & = & -\epsilon^{1/2}\partial_{Y}p\left(Y,T\right),\\
\left(\partial_{y}\phi\left(y,t\right)\right)^{2} & = & \epsilon\left(\partial_{Y}p\left(Y,T\right)\right)^{2},\\
\partial_{y}^{2}\phi\left(y,t\right) & = & -\epsilon\partial_{Y}^{2}p\left(Y,T\right),
\end{eqnarray}
and find
\begin{align}
\partial_{t}\phi & =c-\epsilon\partial_{T}p\nonumber \\
 & =c-\frac{\left\langle \mathbf{W}^{\dagger},\mathcal{D}\mathbf{U}_{c}''\right\rangle }{\left\langle \mathbf{W}^{\dagger},\mathbf{U}_{c}'\right\rangle }\left(\partial_{y}\phi\right)^{2}+\frac{\left\langle \mathbf{W}^{\dagger},\mathcal{D}\mathbf{U}_{c}'\right\rangle }{\left\langle \mathbf{W}^{\dagger},\mathbf{U}_{c}'\right\rangle }\partial_{y}^{2}\phi\nonumber \\
 & -\epsilon\frac{\left\langle \mathbf{W}^{\dagger},\mathbf{f}\right\rangle }{\left\langle \mathbf{W}^{\dagger},\mathbf{U}_{c}'\right\rangle }.
\end{align}
Neumann boundary conditions for $\mathbf{u}\left(x,y,t\right)$ in
$y$-direction, Eq. \eqref{eq:NeumannBoundary}, carry over to Neumann
boundary conditions for $\phi\left(y,t\right).$ Now we exploit the
identity 
\begin{eqnarray}
\frac{\left\langle \mathbf{W}^{\dagger},\mathcal{D}\mathbf{U}_{c}''\right\rangle }{\left\langle \mathbf{W}^{\dagger},\mathbf{U}_{c}'\right\rangle } & = & -\frac{c}{2}.
\end{eqnarray}
which was proven by Kuramoto in Ref. \cite{kuramoto1980instability}.
Finally, we obtain the perturbed nonlinear phase diffusion equation
\begin{align}
\dot{\phi} & =c+\frac{c}{2}\left(\phi'\right)^{2}+\frac{\left\langle \mathbf{W}^{\dagger},\mathcal{D}\mathbf{U}_{c}'\right\rangle }{\left\langle \mathbf{W}^{\dagger},\mathbf{U}_{c}'\right\rangle }\phi''-\epsilon\frac{\left\langle \mathbf{W}^{\dagger},\mathbf{f}\right\rangle }{\left\langle \mathbf{W}^{\dagger},\mathbf{U}_{c}'\right\rangle },
\end{align}
with a perturbation given by
\begin{align}
\left\langle \mathbf{W}^{\dagger},\mathbf{f}\right\rangle  & =\intop_{-\infty}^{\infty}dx\mathbf{W}^{\dagger T}\left(x-\phi\right)\mathbf{f}\left(\mathbf{U}_{c}\left(x-\phi\right),x,y,t\right)\nonumber \\
 & =\intop_{-\infty}^{\infty}dx\mathbf{W}^{\dagger T}\left(x\right)\mathbf{f}\left(\mathbf{U}_{c}\left(x\right),x+\phi,y,t\right),
\end{align}
and the constants
\begin{eqnarray}
\left\langle \mathbf{W}^{\dagger},\mathbf{U}_{c}'\right\rangle  & = & \intop_{-\infty}^{\infty}dx\mathbf{W}^{\dagger T}\left(x\right)\mathbf{U}_{c}'\left(x\right),\\
\left\langle \mathbf{W}^{\dagger},\mathbf{U}_{c}''\right\rangle  & = & \intop_{-\infty}^{\infty}dx\mathbf{W}^{\dagger T}\left(x\right)\mathbf{U}_{c}''\left(x\right).
\end{eqnarray}

\section{\label{sec:FromThePLEEToThePNPDE}From the perturbed linear eikonal
equation to the perturbed nonlinear phase diffusion equation}

We derive the perturbed nonlinear phase diffusion equation \eqref{eq:PerturbedPhaseDiffusionEquation}
from the perturbed linear eikonal equation \eqref{eq:PerturbedLinearEikonalEquation}.
If we parametrize the curve $\boldsymbol{\gamma}$ according to 
\begin{eqnarray}
\boldsymbol{\gamma}\left(y,t\right) & =\left(\begin{array}{c}
\phi\left(y,t\right)\\
y
\end{array}\right),
\end{eqnarray}
the linear eikonal equation becomes 
\begin{eqnarray}
\dfrac{\dot{\phi}}{\sqrt{\left(\phi'\right)^{2}+1}} & =c+\alpha\dfrac{\phi''}{\left(\left(\phi'\right)^{2}+1\right)^{3/2}}.\label{eq:Eikonal0}
\end{eqnarray}
The ansatz for $\phi$ is 
\begin{eqnarray}
\phi\left(y,t\right) & =ct+\Phi\left(Y,T\right)\label{eq:Ansatz0}
\end{eqnarray}
with a slow time scale $T=\epsilon t$ and a stretched space scale
$Y=\epsilon^{1/2}y$. Using Eq. \eqref{eq:Ansatz0} in Eq. \eqref{eq:Eikonal0},
we obtain 
\begin{eqnarray}
\dfrac{c+\epsilon\dot{\Phi}}{\sqrt{\epsilon\left(\Phi'\right)^{2}+1}} & =c+\alpha\epsilon\dfrac{\Phi''}{\left(\epsilon\left(\Phi'\right)^{2}+1\right)^{3/2}}.\label{eq:Eikonal1}
\end{eqnarray}
With $\left(1+\epsilon a\right)^{-1/2}=1-\frac{a}{2}\epsilon+\mathcal{O}\left(\epsilon^{2}\right),$
we expand Eq. \eqref{eq:Eikonal1} and get
\begin{align}
c-\epsilon\frac{c}{2}\left(\Phi'\right)^{2}+\epsilon\dot{\Phi}+\mathcal{O}\left(\epsilon^{2}\right) & =c+\alpha\epsilon\Phi''+\mathcal{O}\left(\epsilon^{2}\right),
\end{align}
or, after some rearrangement,
\begin{align}
\epsilon\dot{\Phi} & =\epsilon\frac{c}{2}\left(\Phi'\right)^{2}+\alpha\epsilon\Phi''+\mathcal{O}\left(\epsilon^{2}\right).
\end{align}
Scaling back to original coordinates $t,\, y$, and introducing the
original function $\phi$ we recover the nonlinear phase diffusion
equation,
\begin{align}
\dot{\phi} & =c+\frac{c}{2}\left(\phi'\right)^{2}+\alpha\phi''.
\end{align}
To treat the perturbation term $\mathbf{f}\left(\boldsymbol{\mathbf{\chi}}\left(\rho,s,t\right),t\right)$,
we express $\boldsymbol{\mathbf{\chi}}$ as
\begin{eqnarray}
\boldsymbol{\mathbf{\chi}}\left(\rho,y,t\right) & = & \boldsymbol{\gamma}\left(y,t\right)+\rho\mathbf{n}\left(y,t\right)\nonumber \\
 & = & \left(\begin{array}{c}
\phi+\rho\left(1+\left(\phi'\right)^{2}\right)^{-1/2}\\
y-\rho\phi'\left(1+\left(\phi'\right)^{2}\right)^{-1/2}
\end{array}\right)\nonumber \\
 & = & \left(\begin{array}{c}
ct+\Phi+\rho\left(1+\epsilon\left(\Phi'\right)^{2}\right)^{-1/2}\\
y-\sqrt{\epsilon}\rho\Phi'\left(1+\epsilon\left(\Phi'\right)^{2}\right)^{-1/2}
\end{array}\right)\nonumber \\
 & = & \left(\begin{array}{c}
ct+\Phi+\rho\\
y
\end{array}\right)+\mathcal{O}\left(\sqrt{\epsilon}\right),
\end{eqnarray}
such that finally
\begin{eqnarray}
\boldsymbol{\mathbf{\chi}}\left(\rho,y,t\right) & =\left(\begin{array}{c}
\rho\\
y
\end{array}\right)+\phi\mathbf{e}_{x}+\mathcal{O}\left(\sqrt{\epsilon}\right).
\end{eqnarray}
The expansion can be truncated after the lowest order in $\epsilon$
because the perturbation $\mathbf{f}$ is also multiplied by $\epsilon$.
Finally we get
\begin{align*}
 & \epsilon\intop_{-\infty}^{\infty}d\rho\mathbf{W}^{\dagger T}\left(\rho\right)\mathcal{G}\left(\mathbf{U}_{c}\left(\rho\right)\right)\mathbf{f}\left(\boldsymbol{\mathbf{\chi}}\left(\rho,s,t\right),t\right)
\end{align*}
\begin{align}
= & \epsilon\intop_{-\infty}^{\infty}d\rho\mathbf{W}^{\dagger T}\left(\rho\right)\mathcal{G}\left(\mathbf{U}_{c}\left(\rho\right)\right)\mathbf{f}\left(\left(\begin{array}{c}
\rho\\
y
\end{array}\right)+\phi\mathbf{e}_{x},t\right)+\mathcal{O}\left(\epsilon^{3/2}\right)\nonumber \\
= & \epsilon\intop_{-\infty}^{\infty}dx\mathbf{W}^{\dagger T}\left(x\right)\mathcal{G}\left(\mathbf{U}_{c}\left(x\right)\right)\mathbf{f}\left(\mathbf{r}+\phi\mathbf{e}_{x},t\right)+\mathcal{O}\left(\epsilon^{3/2}\right),
\end{align}
which is exactly the perturbation term in the nonlinear phase diffusion
equation \eqref{eq:PerturbedPhaseDiffusionEquation}.
\begin{acknowledgments}
We acknowledge financial support from the German Science Foundation
(DFG) within the GRK 1558 (J. L.) and within the framework of the
Collaborative Research Center 910 (S. M. and H. E.).
\end{acknowledgments}
\bibliographystyle{apsrev4-1}
\bibliography{literature}

\end{document}